\documentclass[letter]{aa}

\usepackage[varg]{txfonts}
\usepackage{longtable}
\usepackage{bm}

\newcommand\kms{km~s$^{-1}$}
\newcommand\msun{$M_\odot$}
\newcommand\mhi{$M_{HI}$}

\def\be{\begin{equation}}
\def\ee{\end{equation}}
\def\a40{$\alpha$.40}
\def\arcmin{$^{\prime}$}
\def\arcsec{$^{\prime\prime}$}
\def\dg{$^{\circ}$}

\defcitealias{2013ApJ...768...77A}{A13}
\defcitealias{2013ApJ...777..119F}{F13}

\usepackage{natbib,twoopt}
\usepackage[breaklinks=true]{hyperref} 
\bibpunct{(}{)}{;}{a}{}{,} 
\makeatletter
\newcommandtwoopt{\citeads}[3][][]{\href{http://adsabs.harvard.edu/abs/#3}%
{\def\hyper@linkstart##1##2{}%
\let\hyper@linkend\@empty\citealp[#1][#2]{#3}}}
\newcommandtwoopt{\citepads}[3][][]{\href{http://adsabs.harvard.edu/abs/#3}%
{\def\hyper@linkstart##1##2{}%
\let\hyper@linkend\@empty\citep[#1][#2]{#3}}}
\newcommandtwoopt{\citetads}[3][][]{\href{http://adsabs.harvard.edu/abs/#3}%
{\def\hyper@linkstart##1##2{}%
\let\hyper@linkend\@empty\citet[#1][#2]{#3}}}
\newcommandtwoopt{\citeyearads}[3][][]%
{\href{http://adsabs.harvard.edu/abs/#3}
{\def\hyper@linkstart##1##2{}%
\let\hyper@linkend\@empty\citeyear[#1][#2]{#3}}}
\makeatother

\begin{document}
\title{AGC198606: A gas-bearing dark matter minihalo?}
 \author{E.\ A.\ K.\ Adams
              \inst{1}               \and
               Y. Faerman
               \inst{2}
                 \and
               W.\ F.\ Janesh
               \inst{3} 
               \and
               S. Janowiecki
               \inst{3}
               \and
               T.\ A.\  Oosterloo
               \inst{1,4}
               \and
               K.\  L.\ Rhode
               \inst{3}
               \and
                              R. Giovanelli
               \inst{5}
               \and
               M. P. Haynes
               \inst{5}
                \and
               J. J. Salzer
               \inst{3}
               \and A. Sternberg
               \inst{2}
               \and
               J.\ M.\ Cannon
               \inst{6}
               \and
               R. R. Mu\~noz
               \inst{7}
              }
 \institute{Netherlands Institute for Radio Astronomy (ASTRON), Postbus 2, 7900 AA Dwingeloo, The Netherlands
 		\email{adams@astron.nl}
 		\and
		Raymond and Beverly Sackler School of Physics and Astronomy, Tel Aviv University, Ramat Aviv 69978, Israel
		\and 
		Department of Astronomy, Indiana University, 727 East Third Street, Bloomington, IN 47405, USA		
		\and
		Kapteyn Astronomical Institute, University of Groningen, Postbus 800, 9700 AA Grongingen, The Netherlands
		\and
		Center for Radiophysics and Space Research, Space Sciences Building, Cornell University, Ithaca, NY 14853, USA
		\and
	         Department of Physics and Astronomy, Macalaster College, 1600 Grand Avenue, Saint Paul, MN 55105, USA
		\and
		Departamento de Astronom\'ia, Universidad de Chile, Casilla 36-D, Santiago, Chile}

 
\abstract
{We present neutral hydrogen (HI) imaging observations with the Westerbork Synthesis Radio Telescope
of AGC198606, an HI cloud
discovered in the ALFALFA 21cm survey.
This object is of particular note as it is located 16 \kms\ and $1^{\circ}.2$ from the gas-bearing ultra-faint dwarf galaxy Leo T while having a similar HI linewidth and approximately twice the flux density.
The HI imaging observations
reveal a smooth, undisturbed HI morphology with a full extent of
23\arcmin$\times$16\arcmin\ at the $5\times10^{18}$ atoms cm$^{-2}$ level.
The velocity field of AGC198606 shows ordered motion with a gradient of $\sim$25 \kms\ across $\sim$20\arcmin.
The global velocity dispersion is 9.3 \kms\ with no evidence for a narrow spectral component.
No optical counterpart to AGC198606 is detected.
The distance to AGC198606 is unknown, and we consider several different scenarios:
physical association with Leo T,  a minihalo at a distance of $\sim$150 kpc
based on the models of \citetads{2013ApJ...777..119F}, and a cloud in the Galactic halo.
At a distance of 420 kpc, AGC198606 would have an HI mass of $6.2 \times 10^5$ \msun, an HI radius of 1.4 kpc,
and  a dynamical mass within the HI extent of $1.5 \times 10^8$ \msun.
}

\keywords{galaxies: dwarf --- 
 galaxies: ISM ---
Local Group --- radio lines: galaxies}

\maketitle


\section{Introduction}\label{sec:intro}

While $\Lambda$CDM provides an overall successful theoretical framework towards
understanding the observations of galaxy clustering, properties, and evolution,
large discrepancies have been noted between the predicted number of low 
mass dark matter (DM) halos and the abundance of observed dwarf galaxies \citepads[e.g.,][]{1999ApJ...522...82K}. 
This is generally attributed to the inability of halos to 
retain their baryons, which increases progressively with decreasing halo mass \citepads{2010AdAst2010E...8K}.
The interesting question thus arises: what is the smallest mass halo capable
of hosting an observable baryonic counterpart? 
The discovery of ultra-faint dwarf galaxies (UFDs) around the Milky Way (MW) highlights that galaxies with
minimal stellar components (M$_{star} \lesssim 10^3$ \msun) can exist.
However, it is not clear whether their small stellar populations
are the result of evolution or interaction with the MW \citepads[e.g.,][]{2012AJ....144....4M,2010AJ....140..138M}.

Following the pattern of morphological segregation, gas-rich dwarf systems are 
seen prevalently in the periphery of the Local Group \citepads{1999IAUS..192...17G}.
This pattern continues with the UFDs; 
the most distant UFD (Leo T; d=420 kpc) is the only one with detected HI 
\citepads{2014ApJ...795L...5S}.
In a variation of
the idea previously proposed by \citetads{1999ApJ...514..818B} and \citetads{1999A&A...341..437B}, 
we proposed in \citetads{2010ApJ...708L..22G}  that gas-rich low-mass 
halos may be detected in the HI 21cm line as ultra-compact high velocity clouds
(UCHVCs). 
Using the ALFALFA survey data \citepads{2005AJ....130.2598G}, we presented
a catalog of UCHVCs \citepads{2013ApJ...768...77A} as potential minihalo candidates,
consistent
with the models of \citetads[][hereafter F13]{2013ApJ...777..119F} for gas in low mass DM halos \citepads[see also][]{2002ApJS..143..419S}.
The discovery of the extremely metal-poor, star forming dwarf galaxy Leo P via its HI content in the ALFALFA survey validates the idea that gas-rich galaxies within the Local Group
or its immediate environs can be identified via this tracer \citepads{2013AJ....146...15G,2013AJ....145..149R,2013AJ....146....3S}.

The ALFALFA source AGC198606 has properties quite similar to those of the \citetads{2013ApJ...768...77A}
UCHVCs, except for its radial velocity ($cz_{\odot} = 51$ \kms) being lower than the cutoff used in that work. Its
similarity to the properties of the HI component of Leo T and nearby location (separated
by  $1^{\circ}.2$ and 16 \kms)
was noted by RG and MPH,
and thus it received the nickname "friend of Leo T". 
In this paper we explore the 
possibility that AGC198606 is a system similar to Leo T and Leo P, except for
having a less substantial stellar population.

\begin{figure*}
\centering
\includegraphics[trim=4cm 2cm 4cm 2cm,clip=true,width=\linewidth]{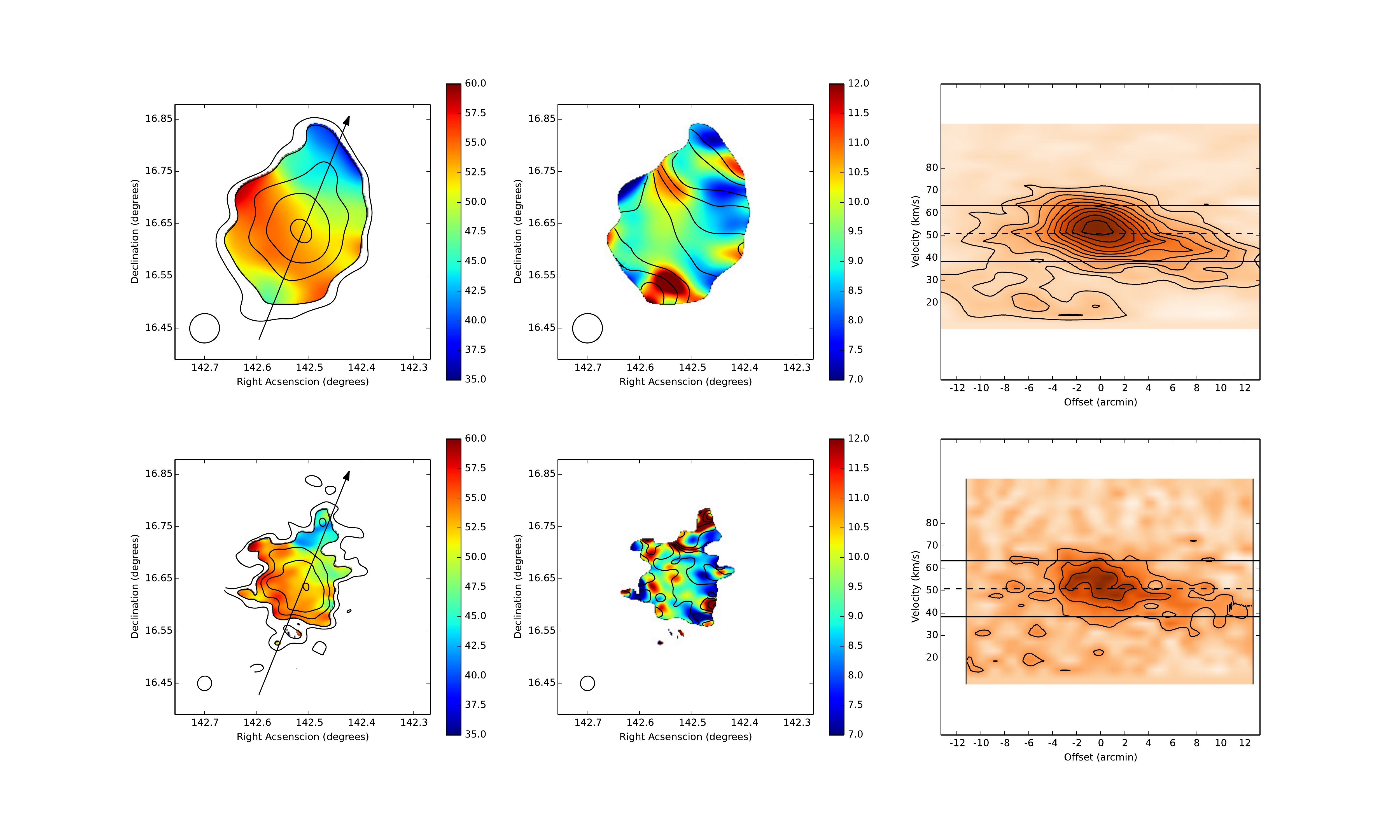}
\caption{HI spatial and kinematic information for AGC198606. 
{\it Upper row}: WSRT data at 210\arcsec\ resolution.
{\it Lower row}: WSRT data at 105\arcsec\ resolution.
{\it Left column}: Velocity field from Gaussian fitting to 4.2 \kms\ spectral data cubes with HI column density
contours overlaid. The contours are [0.5, 0.75, 1.5, 2.5, 4] and [1.25, 1.75, 2.5, 3.5, 5] $\times 10^{19}$ atoms cm$^{-2}$
for the 210\arcsec\ (upper) and 105\arcsec\ (lower) data, respectively.
{\it Middle column}: Velocity dispersion map of AGC198606 from the Gaussian fitting. Contours show lines of constant
velocity from the velocity fields (left) and are at values of [41, 43.5, 46, 48.5, 51, 53.5, 56] \kms.
{\it Right column}: Position-velocity slice along the major axis (shown in left panels). Contours are spaced at 3-$\sigma$; for the 210\arcsec\ and 105\arcsec\ data $\sigma$ is 2.11 and 1.07 mJy beam$^{-1}$, respectively. The dashed line is the recessional velocity $cz=50.9$ \kms, and the solid lines are offset by
$\pm 12.5$ \kms.}
\label{fig:hi}
\end{figure*}


\section{Data}\label{sec:data}

\subsection{WSRT HI Data}
The HI in 
AGC198606 was observed using two 12-hour tracks with the Westerbork Synthesis Radio Telescope (WSRT)
in June 2013 and January 2014.
The spectral setup was a 10 MHz bandwidth divided into 2048 channels, providing a nominal
velocity resolution of 1 \kms.
Standard data calibration and reduction was undertaken in Miriad \citepads{1995ASPC...77..433S}; imaging was done in CASA\footnote{Common Astronomy Software Applications (CASA) is developed and maintained by the National Radio Astronomy Observatory.}. AGC198606 is a low surface brightness object, 
so the data were tapered to lower spatial resolution and binned in velocity to increase the signal-to-noise ratio.
AGC198606 was imaged at 210\arcsec\ resolution (ALFALFA spatial resolution),
105\arcsec\ resolution, and 60\arcsec\ resolution.
At resolutions higher than 60\arcsec, 
the full flux of the system is no longer recovered.

A single channel image ($\Delta v = 46$ \kms) was created for each spatial resolution and used
to isolate the emission;
this was used as an input mask 
for cleaning spectral data cubes with
a velocity resolution of 4.2 \kms.
The observed total flux density is
14.8 $\pm$ 1.5 Jy \kms.
Contours from the total intensity HI maps at 210\arcsec\ and 105\arcsec\ resolution 
are shown in the left column of Figure \ref{fig:hi}. AGC198606 is roughly circular in the inner extent with the outer envelope showing elongation in the north-south direction. 
At  210\arcsec\ resolution, the outermost HI extent  ($5\times10^{18}$ atoms cm$^{-2}$) is 23\arcmin$\times$16\arcmin.
The half-flux radius is 5\arcmin.5 $\pm$ 0\arcmin.5, and
the peak column density at 60\arcsec\ resolution is  $6 \times 10^{19}$ atoms cm$^{-2}$.

The kinematics of AGC198606 were studied by fitting Gaussians to spectral data cubes where the signal
was above the 5-$\sigma$ level.
The resulting velocity fields and position-velocity slices (left and right columns of Figure \ref{fig:hi}, respectively) show
a velocity gradient of $\sim$25 \kms\ across $\sim$20\arcmin, aligned with the major axis of the HI emission.
The turn-over in the velocity gradient at the southern edge of the 210\arcsec\ velocity field
is a result of diffuse emission at low velocities (including minor contamination from Galactic HI, which we estimate to contribute to the final HI flux integral at the $\sim$1\% level) affecting the fit. 
This is also
seen as an increase in the velocity dispersion (upper middle of Figure \ref{fig:hi}).

The velocity dispersion maps 
show the velocity dispersion is at least
7 \kms\ across the full extent of AGC198606.
In order to search for a narrow spectral component from the presence of a cold neutral medium (CNM),
the centers of the fitted Gaussians were used to shift the spectra at each pixel to the recessional velocity of AGC198606. 
The best fit to the resulting global spectrum is a single Gaussian with a velocity dispersion of 9.3 \kms.
 There is no evidence for a multi-phase medium containing a CNM component, 
 although a velocity dispersion as low as 2 \kms\ would be spectrally resolved.

\begin{table}
\caption{AGC198606 Global Parameters}
\label{tab:params}
\centering
\begin{tabular}{c|c}
\hline 
\hline
Property & Value\tablefootmark{a} \\
\hline
R.A.  & 09:30:02.5\\
Decl.& +16:38:08\\
$cz_{\odot}$ & 50.9 \kms \\
$W_{50}$\tablefootmark{b} & 24.7 \kms \\
$\sigma$\tablefootmark{c} & 9.3 \kms \\
$v_{rot}$ & $\sim$14 \kms \\
$S_{HI}$ & $14.8 \pm 1.5$  Jy \kms \\
\mhi & {\bf $3.5\times 10^6 d^{2}_{Mpc}$ }\msun  \\
$\theta_{HI}$\tablefootmark{d} & 11\arcmin $\pm$ 1\arcmin \\
$a\times b$\tablefootmark{e} & 23\arcmin$\times$16\arcmin\\
$r_{HI}$\tablefootmark{f} & 3.3 $d_{Mpc}$ kpc\\
$M_{dyn}$\tablefootmark{g} & $3.5 \times 10^8$ $d_{Mpc}$ \msun\\
$N_{HI}$\tablefootmark{h} & $6\times10^{19}$ atoms cm$^{-2}$\\
\hline
\end{tabular}
\tablefoot{
\tablefoottext{a}{$d_{Mpc}$, the distance in Mpc, parameterizes distance-dependent values}
\tablefoottext{b}{For global spectrum}
\tablefoottext{c}{Intrinsic velocity dispersion after correcting for velocity field}
\tablefoottext{d}{HI half flux angular diameter}
\tablefoottext{e}{HI extent at $5\times10^{18}$ atoms cm$^{-2}$}
\tablefoottext{f}{HI radius at $5\times10^{18}$ atoms cm$^{-2}$}
\tablefoottext{g}{Dynamical mass using $v_{rot}$, $\sigma$, and $r_{HI}$}
\tablefoottext{h}{Peak column density at 60\arcsec\ resolution.}
}
\label{tab:props}
\end{table}

From the WSRT data we can place lower limits on the dynamical mass within the HI extent,
assuming AGC198606 is gravitationally bound.
From \citetads{1996ApJS..105..269H} we adopt: 
\be \label{eq:mdyn}
\mathrm{M_{dyn} = 2.325 \times 10^5 \left( \frac{ V_{rot}^2 + 3 \sigma^2}{km^{2} \, s^{-2}} \right) \left( \frac{r}{kpc} \right) M_{\odot}}
\ee
Assuming there is no rotation and that the global $W_{50}$ represents the dynamics of the system,
we  find a lower limit to the 
dynamical mass of  $2.5\times 10^8 d_{Mpc}$ \msun. 
The alignment of the velocity gradient with the major axis of the HI emission and the clear structure in the 
position-velocity slices (seen in Figure \ref{fig:hi}), are indicative of rotation with an 
uncorrected amplitude of $\sim$12.5 \kms.
The inclination of AGC198606
is estimated from the HI data to be 64\dg;  then the rotation
velocity is  $\sim$14 \kms.
For an intrinsic velocity dispersion of 9.3 \kms, the lower limit
to the dynamical mass then increases to 
$3.5 \times 10^8 d_{Mpc}$ \msun. Varying the rotation velocity
from $[12.5 - 20]$ \kms\ results in a dynamical mass range of $[3.2 - 5.1] \times 10^8 d_{Mpc}$ \msun.

 \subsection{WIYN Data}
 
 Deep observations of AGC198606 were taken on 15 March 2013 with the partially-filled One Degree Imager (pODI;  $\sim$24\arcmin $\times$24\arcmin\ field of view) on the WIYN 3.5m telescope\footnote{The WIYN Observatory is a joint facility of the University of Wisconsin-Madison, Indiana University, the University of Missouri, and the National Optical Astronomy Observatory.} at Kitt Peak National Observatory\footnote{Kitt Peak National Observatory, part of the National Optical Astronomy
Observatory, is operated by the Association of Universities for Research in
Astronomy (AURA) under a cooperative agreement with the National Science
Foundation.}, as part of a larger observing program aimed at characterizing the stellar populations of the UCHVCs (Janesh et al., in prep). 
Nine 300-second exposures were obtained in a dither pattern in both the SDSS $g'$ and $i'$ filters.

The WIYN pODI images were transferred to the ODI Pipeline, Portal, and Archive (ODI-PPA)\footnote{The ODI Pipeline, Portal, and Archive (ODI-PPA) system is a joint development project of the WIYN Consortium, Inc., in partnership with Indiana University's Pervasive Technology Institute (PTI) and with the National Optical Astronomy Observatory Science Data Management (NOAO SDM) Program.} at Indiana University and processed with the QuickReduce data reduction pipeline \citepads{2014ASPC..485..375K}.
The reduced images were reprojected to a common pixel scale, scaled to a common flux level, and combined to create a deep stacked image in each filter. 
SDSS stars present in the images were used to calculate photometric calibration coefficients for converting instrumental magnitudes to calibrated values; errors on the zero points were $<$0.02 magnitude.

The final mean FWHM of point sources is 0.72\arcsec\ in the stacked $g'$ image and 0.78\arcsec\ in the stacked $i'$ image. The 5-$\sigma$ limit on the brightness of a point source in the $g'$-band image is $g' = 25.3$. For the $i'$ image, the corresponding 5-$\sigma$ limit is $i' = 24.6$. We searched for an optical counterpart by first detecting all sources in the images above a modest ($\sim$4-$\sigma$) signal-to-noise threshold, performing photometry on the detected sources, removing extended objects, applying a color-magnitude filter based on the expected stellar population for a range of distances from $0.16-2.5$ Mpc, and implementing a smoothing algorithm to look for an overdensity of stars in the set of filtered objects. The pODI images and our search process reveal no obvious stellar counterpart for AGC198606. 
To estimate an upper limit for the total optical luminosity associated with AGC198606, we masked out bright foreground stars and obvious background galaxies in the combined $i'$-band image.  We then measured the total sky-subtracted flux in an aperture of radius equal to the half-light radius of Leo T \citepads[1.4\arcmin;][]{2007ApJ...656L..13I}, centered on the HI centroid in Table \ref{tab:params}. The measured $i'$-band flux within this aperture yields a limit to the total apparent magnitude of $m_{i'} \simeq 16.5$. At the distance of Leo T (420 kpc), this corresponds to an absolute magnitude of $M_{i'} \simeq -6.6$.  Note that the measured apparent magnitude includes light from stars faint enough to be in a dwarf galaxy at this distance as well as light from any unmasked background galaxies.
Further details of our search methods 
will be presented in Janesh et al. (in prep.).


 \section{The Nature of AGC198606}
 
In this section we discuss two possible scenarios for AGC198606:
that it is a gas-bearing minihalo or part of the population of
HI clouds in the Galactic halo.

 \subsection{Gas-bearing minihalo}

At the distance of Leo T (420 kpc), AGC198606 would have an HI mass of $6.2 \times 10^5$ \msun, an HI radius at the 2$\times 10^{19}$ atoms cm$^{-2}$ level of 600 pc, a full HI extent of 2.8$\times$2.0 kpc, and a dynamical mass 
within the full HI extent 
of $1.5 \times 10^8$ \msun.
Leo T  has a stellar mass of $1.2 \times 10^5$ \msun, an HI mass of 2.8 $\times 10^5$ \msun,
an HI radius of 300 pc (at the $2\times 10^{19}$ atoms cm$^{-2}$ level),
  and an indicative dynamical mass 
  (based on Eqn \ref{eq:mdyn}) of $\sim$1$\times$ $10^7$ \msun\ \citepads{2008MNRAS.384..535R}.
AGC198606 would have about twice the HI mass and size as Leo T, but its
peak HI column density is significantly lower, potentially explaining the apparent lack of a stellar counterpart.
In this scenario,
the two systems would have a projected separation of 8.4 kpc
and could be a bound pair of satellites similar to Leo IV and Leo V \citepads{2010ApJ...710.1664D}.
The ALFALFA data do not reveal HI emission connecting AGC198606 to Leo T; 
low level emission connecting the two cannot be ruled out as Leo T lies
at the same velocity as strong foreground Galactic HI emission.

Applying the models from \citetalias{2013ApJ...777..119F} 
for a flux density of 14.8 Jy \kms, a half-flux radius of 5\arcmin.5,
and a peak column density of $6\times10^{19}$ atoms cm$^{-2}$ gives a distance estimate of $120-180$ kpc  for a typical flat-cored halo.
At a distance of 150 kpc, AGC198606 would have an HI mass of $7.9 \times 10^4$ \msun, an HI half-flux radius of 240 pc, 
a full HI extent of $1.0\times0.70$ kpc, and a dynamical mass of $5.3 \times 10^7$ \msun.
These HI properties are similar to those of Leo T, and the HI kinematics are similar to Leo P.
HI imaging observations of Leo P reveal a rotational velocity of 15 \kms\ at an HI radius of 500 pc
\citepads{2014AJ....148...35B}.
With less HI than Leo T and no CNM (seen in both Leo T and Leo P), the lack of a clear stellar component is understandable.
However, at this close distance, the question arises of how a small object
could retain its HI gas in the presence of the MW's hot corona.

\begin{figure}
\centering
\includegraphics[keepaspectratio,width=\linewidth]{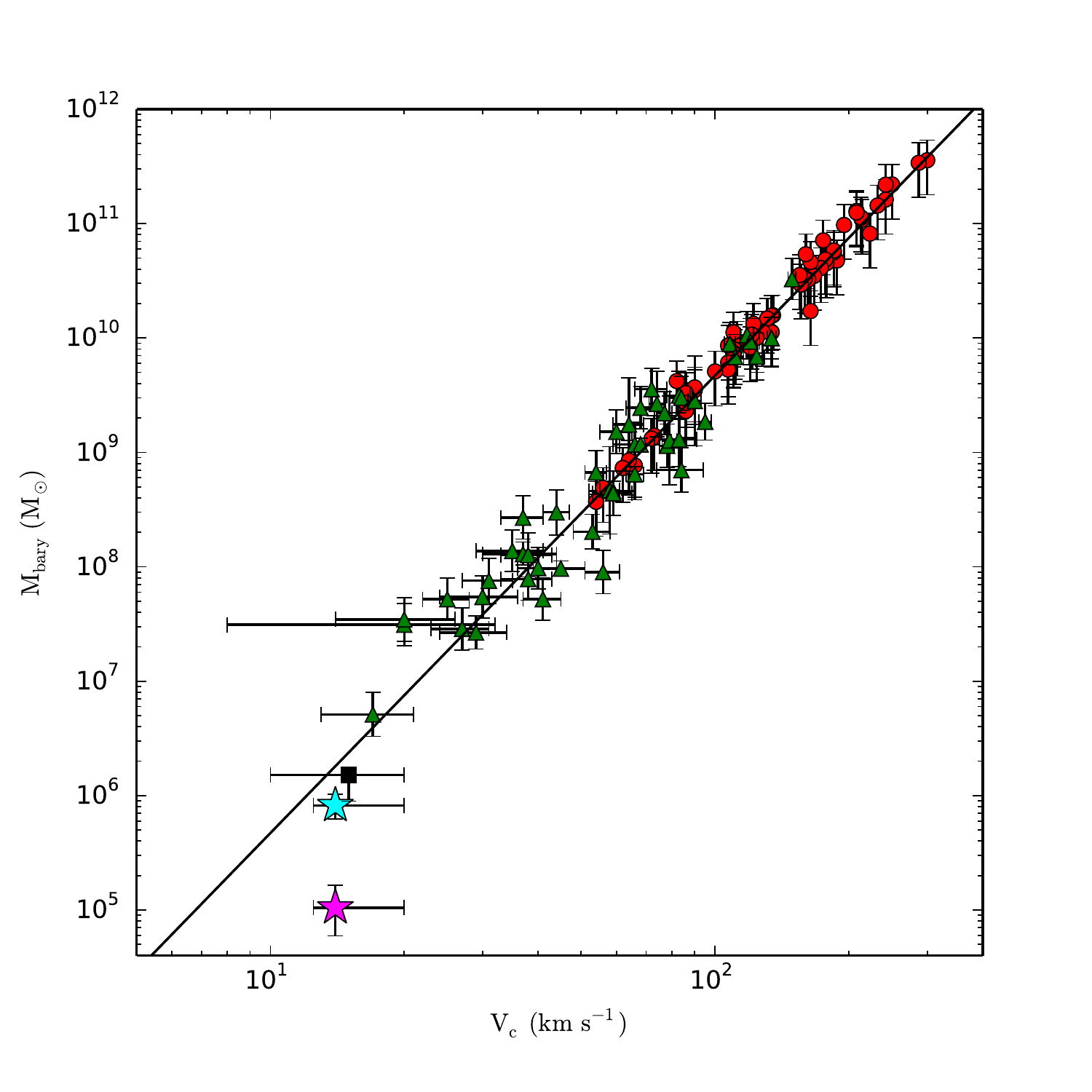}
\caption{The baryonic Tully-Fisher relation from \citetads[][black line]{2012AJ....143...40M} shown with the data points of \citetads[][green triangles]{2012AJ....143...40M} and \citetads[][red circles]{2005ApJ...632..859M}. 
Leo P is shown as the black square \citepads{2014AJ....148...35B}. AGC198606 is shown  at 420 kpc (cyan star) and 150 kpc (magenta star). }
\label{fig:btf}
\end{figure}

 The baryonic Tully-Fisher relation (BTFR) relates baryonic mass to rotational velocity
 over 5 orders of magnitude \citepads{2012AJ....143...40M}.
 Figure \ref{fig:btf} shows the BTFR from \citetads{2012AJ....143...40M} with AGC198606 overplotted 
 based on its total atomic gas mass ($1.33\times M_{HI}$ to account for helium) 
 for the two distances considered above. In both cases, it lies below the canonical BTFR,
 although the Leo T scenario is consistent with the scatter at low velocities.
In order for AGC198606 to fall on the canonical BTFR based solely on its atomic gas mass,
 it would need to be at a distance of
620 kpc ($M_{gas} =  1.8 \times 10^6$ \msun).
At this distance, it would have an HI diameter of 4.1 kpc, $\sim$4 times greater than
expected for its HI mass \citepads{2008MNRAS.386.1667B}; the lack
of a stellar counterpart for such a system would be exceptionally rare.

\subsection{Galactic halo HI cloud}

 Alternatively, AGC198606 could be a nearby gas cloud associated with the Galactic halo.
  For a representative distance of $\sim$10 kpc, it would have
 an HI mass of 350 \msun\ and an HI size of $67\times47$ pc.
  High resolution HI observations of compact Galactic halo clouds  generally reveal an irregular morphology, often consisting of several small cores with narrow velocity widths at
 scales of a few arcminutes \citepads[e.g.,]{2002A&A...391...67D,2004A&A...426L...9B}. 
 In contrast, the HI morphology of AGC198606 remains rather smooth and undisturbed down to the imaging limits,
 and there is no evidence for cold condensations of gas.
While we cannot rule this scenario out, the HI morphology and kinematics of AGC198606 are not a good match to previous observations of Galactic halo clouds.


\section{Summary}
 AGC198606 is a compact HI cloud identified within the ALFALFA HI survey. It is of particular interest
 as it is near the gas-bearing UFD Leo T  spatially and kinematically,
 and it has similar HI
 properties as measured by the single-dish ALFALFA survey. 
 This makes it an excellent candidate to represent a (nearly) starless gas-bearing DM halo.
  Imaging observations with WSRT show that AGC198606 has an HI mass of $3.5\times10^6 d^{2}_{Mpc}$ \msun,
  an HI radius of 3.3 $d_{Mpc}$ kpc,
 and an ordered velocity gradient of $\sim$25 \kms\ along the HI major axis.
 Deep optical imaging with the WIYN 3.5m telescope reveal no obvious stellar counterpart; future
 work will quantify the stellar population that could be detected in these images (Janesh et al. in prep).
  Without
 the direct identification of a stellar counterpart to constrain the distance to AGC198606, its true nature
 remains uncertain, and
we discuss several plausible distances for AGC198606.
If physically associated with Leo T, AGC198606 would have about twice the HI mass and size.
 Alternatively, it could be at a closer distance  of 150 kpc suggested by the  models in \citetalias{2013ApJ...777..119F}.
 The HI morphology of AGC198606 is different from observations of clouds in the Galactic halo 
with no evidence for cores, although they could exist at scales smaller than 60\arcsec.
HI imaging observations of other UCHVCs 
will offer further context for the UCHVCs and help determine if the HI structure of AGC198606
is 
different from that of Galactic halo clouds.


\begin{acknowledgements}
We thank the anonymous referee for valuable input.
The Westerbork Synthesis Radio Telescope is operated by the ASTRON (Netherlands Institute for Radio Astronomy) with support from the Netherlands Foundation for Scientific Research (NWO). The ALFALFA work at Cornell is supported by NSF grants AST-0607007 and AST-1107390 to R.G. and M.P.H. and by grants from the Brinson Foundation. 
K.L.R. and W.F.J. acknowledge support from NSF CAREER award AST-0847109. J.M.C. is supported by NSF grant AST-1211683.
R.~R.~M.~acknowledges partial support from CONICYT Anillo project ACT-1122 and project BASAL PFB-$06$ 
as well as FONDECYT project N$^{\circ}1120013$.
\end{acknowledgements}


\bibliographystyle{aa}
\bibliography{refs}
\defcitealias{2013ApJ...777..119F}{F13}


\end{document}